\documentclass[a4paper,12pt]{article}
\usepackage{indentfirst}
\usepackage[numbers,sort&compress]{natbib}
\pdfoutput=1 
\usepackage[utf8]{inputenc}
\usepackage{pgf}
\usepackage{epsfig}
\usepackage{amsmath}
\usepackage{amssymb} 
\usepackage{amsfonts}
\usepackage{color}
\numberwithin{equation}{section} 
\usepackage[width=17cm]{geometry} 

\newcommand\beqa{\begin{eqnarray}}
\newcommand\eeqa{\end{eqnarray}}
\newcommand\beq{\begin{equation}}
\newcommand\eeq{\end{equation}}
\newcommand\beal{\begin{align} }
\newcommand\eeal{\end{align} }
\newcommand\nn{\nonumber}
\def\bi{\bibitem}
\def\bit{\begin{itemize}}
\def\eit{\end{itemize}}
\def\nn{\nonumber}
\def\f{\frac}

\def\[{\left[}
\def\]{\right]}
\def\({\left(}
\def\){\right)}

\def\..{\left.}
\def\.{\right.}
\def\tl{\tilde}

\usepackage{setspace}
 
\begin{document}
 
\Large

\title{\textbf{ Ultra High Energy Cosmic Rays in light of the Lorentz Invariance Violation Effects within the Proton Sector }}

\author{Guo-Li Liu\footnote{Corresponding author, e-mail: guoliliu@zzu.edu.cn}, Xinbo Su, Fei Wang \footnote{Corresponding author, e-mail: feiwang@zzu.edu.cn} \\
School of Physics, Zhengzhou University, Zhengzhou 450001, China}
%
\maketitle

\abstract{ 
Tiny Lorentz Invariance Violation (LIV) effects, potentially arising from quantum gravity-induced spacetime structures, may also manifest in the proton sector, offering a plausible pathway to test Planck-scale physics through high-energy cosmic phenomena.
Our analysis reveals that even minuscule LIV effects in the proton sector can significantly elevate the photon threshold energy for photopion production to $\cal {O}$
(0.1  to $10^3$ eV),
orders of magnitude higher than in Lorentz-symmetric scenarios. Consequently, protons in ultra-high-energy cosmic rays (UHECRs) can propagate for very long distances without significant energy loss via photopion interactions with cosmic microwave background (CMB) photons. This suppression of attenuation may provide a plausible explanation for the observed cosmic-ray events exceeding the Greisen-Zatsepin-Kuzmin (GZK) cutoff energy.
We further demonstrate that when both leading-order and next-to-leading-order LIV effects are considered, higher-order LIV contributions induce discontinuous transitions in the GZK cutoff energy spectrum. Observations of proton-dominated UHECRs beyond the GZK threshold could provide constraints the LIV energy scale.
Offering insights into the ultraviolet regime of LIV theories near the Planck scale, UHECRs may serve as a sensitive probe of LIV and provide a means to test quantum gravity predictions by constraining deviations from Lorentz symmetry in extreme-energy regimes.
}
\vspace{2cm}
\newpage
\tableofcontents  

\section{Introduction}
Ultra-high energy cosmic rays (UHECRs), which are most likely of extraterrestrial origin \cite{Aab:2017tyv,Abreu:2012ybu,Tinyakov:2015qfz,Abbasi:2016kgr,Blasi:2014roa,ppnp125-2022-103948}, are primarily composed of protons and other atomic nuclei with energies exceeding 1 EeV. Cosmic rays with energies above $\sim5 \times 10^{19} \text{eV}$ have a finite propagation range due to energy losses caused by interactions with the cosmic microwave background (CMB). 
Specifically, cosmic ray protons with energies of $10^{20}$ eV lose most of their energy through photopion production over long distances. This results in a finite mean free path of approximately 100 Mpc for protons with energies of $10^{20}$ eV, and the distance could be even shorter for higher energies or heavier nuclei \cite{2110.09900}. As a consequence, UHECRs originating from distant sources are rarely detected on Earth. This theoretical upper limit on the energy of UHECRs is known as the Greisen-Zatsepin-Kuzmin (GZK) cutoff \cite{Greisen,Zatsepin}. 
\footnote{Despite the presence of the GZK cutoff, cosmic rays with energies as high as $10^{20}$ eV can still be detected. This phenomenon can be attributed to several factors, including nearby sources\cite{Greisen, Zatsepin} , the composition of heavy nuclei such as iron\cite{Allard-2007, Kotera-Olinto-2011} , special sources like active galactic nuclei and gamma-ray bursts \cite{Hillas-1984, Waxman-1995}, the statistical nature of observations due to fluctuations or unique propagation paths \cite{Abreu-2011, Batista-2016}, and the spectral extension, which indicates that the GZK cutoff is not an absolute limit but rather a characteristic inflection point on the energy spectrum \cite{Berezinsky-Grigorieva-1988, Blasi-2013}. Certainly, as stated in this article, the potential explanations for these phenomena may also involve new physics \cite{Anchordoqui-2014, Berezinsky-1997, Sigl-prd-2004}.}


 Around the same time the GZK cutoff was proposed, the Volcano Ranch experiment \cite{1963-1st-uhecr-event} and the SUGAR array \cite{1968-2nd-uhecr-event} reported the first published events with energies exceeding $10^{20}$ eV.
Subsequently, it seems that other events with energy scale above GZK cutoff emerged ~\cite{1986-3nd-uhecr-event,uhecr-4,uhecr-5,uhecr-6,uhecr-7,uhecr-8,uhecr-9}. 
High-energy events with comparable energies have also been observed in recent studies ~\cite{AGASA-1,AGASA-2,uhecr-6,Hires-2,pao-1,pao-2,TA-1}, as well as in the Pierre Auger Observatory and Telescope Array\cite{JCAP052023_024,ICRC2021_338}.
\footnote{One must note that AGASA\cite{AGASA-1, uhecr-7,Uchihori}, Yakutsk\cite{Ivanov}, HiRes\cite{Abbasi}, Auger\cite{Abreu-2011}, and Telescope Array\cite{Tokuno}
conducted observations of UHECRs under different time periods and technical conditions. As a result, the physical significance differs in several aspects, primarily including energy spectrum and observation of the GZK cutoff, anisotropy and source localization, composition analysis, composition analysis experimental techniques and systematic errors. In summary, AGASA and Yakutsk initially challenged the GZK cutoff and hinted at UHECR clustering, but their results were limited by experimental uncertainties. In contrast, HiRes, Auger, and Telescope Array provided stronger evidence for the GZK cutoff and more precise data on anisotropy and composition. These differences highlight the impact of technological progress and offer key insights into the origin, acceleration, and propagation of UHECRs.}

%
Although the sources of UHECRs are still unclear, we suggest that their propagation can possibly involve some effects from new physics, such as perturbative Lorentz Invariance Violation (LIV), which acts as the effective description of some quantum gravity theory at low energies. 
Even very tiny LIV effects, potentially arising from some ultraviolet (UV) theory, can manifest in the low-energy Standard Model (SM). These effects, though minuscule, might explain certain peculiar features observed in the physics of UHECRs, such as an extended opacity horizon for UHECRs~\cite{Coleman, Stecker, Scully, TorriUHECR, TorriPhD},  because such gradual effects can increase with energy. 
In the following, we will propose a potential explanation for some UHECR events observed above the GZK energy threshold within this framework, without implying that all of these events require new mechanisms.


The composition of UHECRs is believed to be roughly categorized into lighter nuclei, (e.g., Protons, helium nuclei) and heavier nuclei (e.g., Carbon, Nitrogen, Oxygen, Silicon, Iron). 
Among the components of UHECRs, protons are relatively abundant,  although their abundance decreases at the highest energies, as suggested by recent data from the Pierre Auger Observatory\cite{2406.06315}.
And they may only be deflected slightly by magnetic fields and follow relatively direct trajectories. 
\footnote{Since the Lorentz force is given by $ F = qvB = \frac{mv^2}{r} $, the radius of deflection $ r $ can be expressed as $ r = \frac{mv}{qB} = \frac{p}{qB} $, where $ p $ is the momentum and $ q $ is the electric charge. This means the deflection radius is determined by the momentum $ p $ and the charge $ q $. For particles with nearly the same momenta, those with lower electric charges will experience smaller deflection radii. Consequently, protons and heavy nuclei with energies of the same order will behave differently: protons, having a lower charge, will exhibit smaller deflections compared to heavy nuclei.}
In this paper, we will explore potential effects of LIV in the proton sector on the propagation of UHECRs.

Numerous theoretical models of LIV have been proposed within various new physics frameworks~\cite{9809521-Koste1, Coleman, 0601236-vsr-Cohen, AmelinoCamelia, AmelinoCamelia2, AmelinoCamelia3, AmelinoCamelia4,string,Chengyi2021,Chengyi2022,Chengyi2023,Rovelli2008, Ashtekar2021, lihao2023,DSR1-1, DSR1-2, 0112090-DSR2-1-Smolin2, DSR2-2}.
%
In our analysis, we adopt a straightforward and minimal extension to the Standard Model (SM) of particle physics~\cite{TorriHMSR, 2110.09900}.
The inclusion of LIV can lead to slight modifications in the relevant kinematics and alter the allowed phase space for processes such as photopion production.

In the following sections, we will focus on the LIV for proton sector incorporating a modified dispersion relation from several perspectives:
the geometry of spacetime, the extension of the SM, and the altered kinematics, etc.
 Possible LIV effects for interactions between UHECRs and CMB will be explored. This work is organized as follows.
In Section \ref{sec:LIV}, we introduce a specific form of LIV for the proton sector and discuss its possible origins.  
In Section \ref{sec:propaga-cal}, we explore the phenomenological implications of LIV on the interactions between UHECRs and the CMB. These effects can influence the propagation of UHECRs and provide insights into explaining events beyond the GZK cutoff.  
In Section \ref{sec:conclusions}, we outline future directions for extending this analysis to include a more realistic cosmic ray composition scenario.


\section{Lorentz Invariance Violation of Protons}\label{sec:LIV}

Lorentz invariance plays a key role in Einstein's special relativity and serves as a cornerstone for both general relativity and quantum field theory in modern physics.
 However, Lorentz symmetry
may not be exact near the Planck scale ($M_\text{Pl}= 1.22\times 10^{19}$ GeV),  possibly because of the modification of the dispersion relation. 
From a bottom-up point of view, such a modification can effectively describe an LIV scheme that modifies the phenomenology of particle physics
and at the same time preserves space-time isotropy and does not introduce exotic interactions. It may origin from geometrized  interaction with the assumed background quantum structure, akin to the approach taken in the Homogeneously Modified Special Relativity (HMSR) theoretical framework \cite{TorriHMSR, 2304.12767, 9809521-Koste1, Coleman, 0601236-vsr-Cohen, AmelinoCamelia, AmelinoCamelia2, AmelinoCamelia3, AmelinoCamelia4, Smolin1, 0112090-DSR2-1-Smolin2}.


Conventional dispersion relations can be modified by the introduction of various LIV terms that are suppressed by a factor of
($E/M_\text{Pl})^n$, where the exponent $n$ is sensitive to the underlying UV theories.
The modified dispersion relations (MDRs) for massive fermions can be expressed as follows \cite{1406.4568}.
\beqa
E_p^2 =m_p^2+|\vec{p}|^2\left[ 1 -\eta_n (\frac{|\vec{p}|}{E_\text{LV,n}})^n\right].
\label{LIV-n-order}
\eeqa
The four-momentum of the massive fermion, such as protons, is denoted by ($E_p,~\vec{p}$), with
$|\vec{p}|$ the magnitude of the three-momentum vector.
The exponent $n$ in the Lorentz-violating term determines the energy dependence:
$n=1$ corresponds to linear energy dependence, and
$n=2$ corresponds to quadratic energy dependence.
The sign of the Lorentz violation correction is given by $\eta_n=\pm 1$, where
$\eta_n=+ 1$ indicates that 
the maximum attainable velocity of this massive particle is less than convential vacuum light speed (subluminal case), 
and
$\eta_n=- 1$ indicates the opposite (superluminal) case.
The suppression scale of the $n$-th order Lorentz violation is characterized by the Lorentz violation parameter
$E_\text{LV,n}$.
A modified dispersion relation, such as Eq.~(\ref{LIV-n-order}),
can possibly affect significantly the propagation of cosmic particles with ultra
high energies.
It is important to note that, in practice, each particle species is assumed to have its
own unique dispersion relation \cite{0812.0121, 2304.12767, 2105.07967, 2110.09900}. 

If Lorentz symmetry is violated in the proton sector while being preserved for other particles, cosmic ray protons of varying energies scattering with cosmic microwave background (CMB) photons would experience differing time delays ($\Delta t$) upon reaching Earth. Given the vast cosmological propagation distances of cosmic rays and the extremely high energies of these protons, the time delay ($\Delta t$) induced by Lorentz invariance violation (LIV) could, to some extent, be measurable. However, other factors, such as magnetic field deflections, may also contribute significantly to the observed delay.
\footnote{In fact, the time delay of UHECR protons may arise not only from Lorentz symmetry violation but also from other sources, such as magnetic field deflections and interactions with the propagation medium \cite{sigl2004,taylor2011,anchordoqui2001,berezinsky1990,kotera2011,watson2013}. Therefore, attributing the observed time delay solely to Lorentz symmetry violation and claiming its measurability lacks sufficient rigor.}
This is analogous to the case of high-energy photons, which, under modified dispersion relations, demonstrate vacuum dispersion effects~\cite{1406.4568}.

\subsection{The MDR arising from 
 LIV}\label{sec:mody-SR}

%
In Eq. (\ref{LIV-n-order}), the perturbation term, which may vary for different particle species, is expressed as the function  
$f\left(\frac{p}{E_\text{LV,n}}\right) = f\left(\frac{|\vec{p}|}{E_\text{LV,n}}\right) = \left(\frac{|\vec{p}|}{E_\text{LV,n}}\right)^n$.  
To maintain rotational invariance, this function depends exclusively on the magnitude of the three-momentum vector $|\vec{p}|$. 
The modified dispersion relation (MDR), defined as the square of the four-momentum combined with this perturbative term, is given by
\beqa  \nn
MDR\Rightarrow F^2(p)&\equiv& p^2+\eta_n |\vec{p}|^2 (\frac{|\vec{p}|}{E_\text{LV,n}})^n\equiv m^2\\ \nn
&=&E^2-|\vec p|^2 + \eta_n |\vec{p}|^2 (\frac{|\vec{p}|}{E_\text{LV,n}})^n~.
\label{mdr-n-order}
\eeqa
Such a $F(p)$ can  be a candidate for Finsler-Lagrange function (Finsler pseudo-norm), in which
the very small term $(\frac{|\vec{p}|} {E_\text{LV,n}})^n$ 
with
\beqa\label{a2}
\frac{|\vec p|}{E_\text{LV,n}}\rightarrow\epsilon\ll 1,
\eeqa
can be seen as a perturbation and integrated into the framework of pseudo-Finsler geometry~\cite{Pfeifer1, Pfeifer2, Pfeifer3, Javaloyes, Bernal}.
%
Each massive particle specie can have its own MDR (of the form $F^2(p)$) and consequently its own distinct maximum velocity $(1-\epsilon^n)$, which is similar to the case in Very Special Relativity, within which every massive particle admits a different personal maximum attainable velocity~\cite{Coleman}
\begin{equation}
c'=\left(1-\frac{|\vec p|}{E_\text{LV}}\right)\rightarrow(1-\epsilon).
\end{equation}


\subsection{The Modified Metric  }

Through a Legendre transformation that expresses $p_\mu$ in terms of $\dot{x}^\mu$ up to leading order,  
\beqa  
p_\mu = \frac{g_{\mu\nu} \dot{x}^\mu}{\sqrt{g_{\mu\nu}(x, \dot{x}) \, \dot{x}^\mu \dot{x}^\nu}},  
\eeqa  
and utilizing the relation for the L-metric,
\begin{equation}  
\label{a9}  
\widetilde{g}^{\mu\alpha} \, g_{\alpha\nu} = \delta^{\mu}_{\,\nu},  
\end{equation}  
we can express the Hamiltonian as  
\begin{equation}  
\label{a12}  
\mathcal{H} = \vec{p} \cdot \dot{\vec{x}} - \mathcal{L} \simeq \sqrt{\widetilde{g}^{\mu\nu}(p) \, p_{\mu} \, p_{\nu}} = F(p),  
\end{equation}  
which is derived from the action of a massive particle in terms of proper time,  
\beqa  
S = -m \int dl = -m \int d\tau \left[ g_{\mu\nu}(x, \dot{x}) \, \dot{x}^\mu \dot{x}^\nu \right]^{1/2}.  
\eeqa   

The function $ F^2(p) $ can serve as the Finsler-Lagrange function (or Finsler pseudo-norm) in momentum space, with the corresponding L-metric given by:
\beqa
\widetilde{g}(p)^{\mu\nu}&\equiv& \frac{1}{2}\frac{\partial}{\partial p^{\mu}}\frac{\partial}{\partial p^{\nu}} F^2(p) \nn\\
&=&\frac{1}{2}\frac{\partial}{\partial p^{\mu}}\frac{\partial}{\partial p^{\nu}}\left(p^2(E,\,\vec{p})+ \eta_n|\vec{p}|^2 \left(\frac{|\vec{p}|}{E_\text{LV,n}}\right)^n\right),\nn\\
&=&\left(
           \begin{array}{cc}
             1 & \vec{0} \\
             \vec{0}^{t} & -\left[1+\eta_n\left(\frac{|\vec{p}|}{E_\text{LV,n}}\right)^n\right]\mathbb{I}_{3\times3} \\
           \end{array}
      \right).
\label{aa4}
\eeqa
The non-diagonal components do not contribute to the computation of the MDR and can therefore be neglected. Furthermore, the resulting L-metric in coordinate space can be determined as
\begin{equation}
\label{a6}
g(x,\,\dot{x}(p))_{\mu\nu}=\left(
                     \begin{array}{cc}
                         1 & \vec{0} \\
                         \vec{0}^{t} & -\left[1-\eta_n\left(\frac{|\vec{p}|}{E_\text{LV,n}}\right)^n\right]\mathbb{I}_{3\times3} \\
                     \end{array}
                  \right),
\end{equation}
after employing the Legendre transform and neglecting the derivatives with respect to momentum \cite{TorriHMSR}.

The generalized vierbein, defined by
\begin{equation}
\begin{split}
\label{a13}
&g_{\mu\nu}(\dot{x})=e_{\mu}^{\,a}(p(\dot{x}))\,\eta_\text{ab}\,e_{\nu}^{\,b}(p(\dot{x}))\\
&\widetilde{g}^{\mu\nu}(p)=e^{\mu}_{\,a}(p)\,\eta^{ab}\,e^{\nu}_{\,b}(p)~,
\end{split}
\end{equation}
can be obtained from the L-metrics from Eq.(\ref{aa4}) and Eq.(\ref{a6})
\begin{equation}
\label{a14}
\begin{split}
& e^{\mu}_{\,a}(p)=        \left(
                             \begin{array}{cc}
                               1 & \vec{0} \\
                               \vec{0}^{t} & \sqrt{1-\eta_nf\left(\left(\left|\vec{p}\right|/E\right)\right)}\,\mathbb{I}_{3\times3} \\
                             \end{array}
                             \right)
                 =        \left(
                             \begin{array}{cc}
                               1 & \vec{0} \\
                               \vec{0}^{t} & \sqrt{1-\eta_n\left(\frac{|\vec{p}|}{E_\text{LV,n}}\right)^n}\,\mathbb{I}_{3\times3} \\
                             \end{array}
                           \right)\\ \\
& e_{\mu}^{\,a}(p)=        \left(
                             \begin{array}{cc}
                               1 & \vec{0} \\
                               \vec{0}^{t} & \sqrt{1+\eta_nf\left(\left(\left|\vec{p}\right|/E\right)\right)}\,\mathbb{I}_{3\times3} \\
                             \end{array}
                              \right)
                 =        \left(
                             \begin{array}{cc}
                               1 & \vec{0} \\
                               \vec{0}^{t} & \sqrt{1+\eta_n\left(\frac{|\vec{p}|}{E_\text{LV,n}}\right)^n}\,\mathbb{I}_{3\times3} \\
                             \end{array}
                           \right)
\end{split}
\end{equation}
As each particle resides in its unique momentum-dependent modified pseudo-Finsler space-time, all its relevant physical quantities can explicitly depend on the momentum. It can be obtained that all components of the affine connection, derived from the L-metric, vanishes identically or vanishes at the leading order (after neglecting the derivatives of the 0 degree homogeneous perturbation function). The spin connection, which is needed for spinors in curved space-time, is also negligible at the leading order, since every element is again proportional to derivatives of perturbation function $f(p)$. Consequently, the resulting geometric structure is an asymptotically flat pseudo-Finsler structure. With $\Gamma^{\mu}=e^{\;\mu}_{a}(p)\,\gamma^{a}$, it can be proven with modified Clifford algebra that MDR holds for the spinor fields.


\section{LIV Effects on UHECRs Propagation}\label{sec:propaga-cal}
UHECRs experience various attenuation processes as they interact with background photons. These processes are significantly influenced by the specific composition and energy levels of the UHECRs.
\beqa
\label{pr1} &&
A+\gamma\,\rightarrow\,(A-1)+n, ~~~ {\rm photo-dissociation}\\\label{pr2}&&
p+\gamma\,\rightarrow\,p+e^{-}+e^{+},~~~ {\rm  e^{\pm}~pair~ production}\\\label{pr3}&&
p+\gamma\,\rightarrow\,\Delta\,\rightarrow\,p+\pi^{0},~~\ {\rm photopion ~process}  \\\label{pr4}&&
p+\gamma\,\rightarrow\,\Delta\,\rightarrow\,n+\pi^{+}, ~{\rm photopion ~process}.
\eeqa
When protons have energies below the threshold $E_\text{p,th}=5\times10^{19}\,\text{eV}$,
the production of electron-positron pairs on the CMB, as described in Eq.~(\ref{pr2}), is the dominant loss mechanism for protons,
while for protons with even high energies, the processes outlined in Eqs.~(\ref{pr3})(\ref{pr4})
become the primary modes of attenuation\cite{pp-2000-op}.
The photopion production processes are the central mechanism that is responsible for the GZK cutoff phenomenon of protons,  establishing an upper limit on the energy of UHECRs originating from distant sources.


%
By analyzing the interactions between photons and protons, we can discern the characteristics of the GZK effect and revisit the propagation of high-energy protons across the Universe.
Within the framework of the SM of particle physics,
the collision of a proton with a photon can proceed via an intermediate  resonant $\Delta$ state,
which can be either real or virtual.
Such dominant processes can be depicted by $p\gamma\to \Delta\to N\pi^+/p\pi^0$.
Therefore, to produce resonantly a real $\Delta$ particle, the center-of-mass energy in the collision
 between the proton and the CMB photon must exceed the $\Delta$ rest mass.~\footnote{
Should the kinematic prerequisites for generating a real $\Delta$ particle not be satisfied,
the incorporation of LIV could significantly reduce the photopion production phenomenon.
This, in turn, would have a consequential impact on the GZK cutoff.}.

The interaction between high-energy protons and the CMB background photons significantly impedes the ability of these protons to travel vast distances across the Universe.
Denoting $E_p$ the energy of the high-energy proton and $\epsilon_b$ the energy of the CMB photon,
the principles of special relativity allow us to deduce the threshold condition for this process
\beq
\epsilon_{b} = \frac{(m_{\pi} + m_{N})^2 - m_{p}^2}{4E_{p}},
\label{org-eps}
\eeq
where $m_p$, $m_{N}$, and $m_\pi$ denote the masses of protons, neutrons, and pions, respectively.
 Given the average energy $\epsilon_b\approx6.35\times10^{-4}$ eV for CMB photons,
the threshold energy for proton can be calculated by Eq. (\ref{org-eps}) to be 
 $E_\text{p,th}= 5\times 10^{19}$ eV.
Protons with energies exceeding this threshold are unlikely to reach Earth due to such photopion processes involving the CMB photons. It can be seen from  Eq. (\ref{org-eps}) that the threshold energy for proton decreases as the energy of the background photon increases.
As a rough estimation, for $\epsilon_{b}$ from $10^{-3}$ eV to $1$ eV, the resulting thresholds
$E_\text{p,th}$ would span from $10^{19}$ eV to $10^{17}$ eV.

\begin{figure}[htb]
\centering
\includegraphics[scale=1.0]{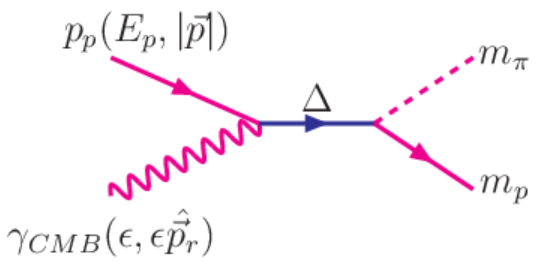}
\caption{\textit{photopion process, through $\Delta$ particle resonance}.}
\label{fig1:rp-pn}
\end{figure}

We anticipate that  LIV effects can only result in minor deviations from ordinary GZK phenomenon.
Given that the dispersion relation is altered by LIV as depicted in Eq. (\ref{LIV-n-order}), 
the kinematics of the reaction $p\gamma\to N X$ should be modified accordingly, leading to a modification of the allowed proton energy spectrum in UHECRs, where $N$ represents nuclei and
$X$ denotes mesons.
To simplify the analysis, we will focus on the process $p\gamma\to N \pi$
in the following discussion (with the relevant Feynman diagram shown in Fig. \ref{fig1:rp-pn}),
where $\hat{\vec{p}}_r$ denotes the unit vector of the photon momentum.

\begin{itemize}

\item  LIV only in the proton sector:

\begin{itemize}
\item Only $n=1$ order LIV term is present:

 As noted previously, each particle species can practically have its own unique dispersion relation. For simplicity, in the process $p\gamma\to N \pi$, we assume that LIV is present only in the proton sector, while other sectors remain Lorentz invariant.~\footnote{More practical discussions would consider that both fermions and mesons are LIV, such as that in Ref.\cite{2404.15838}.
However, it is also feasible to consider only the LIV from the initial proton, since in the process $p\gamma  \to N\pi $, the initial particle energy is much higher, which is more accordance with the assumption that LIV occures at very high energies. More  importantly, as seen later, if the proton is LIV, it will require an increase of the minimum energy of the photons in the process $p\gamma\to N \pi$, much higher than the CMB photon energy. In this case, it would be quite difficult to have this process.}.
 The dispersion relation for protons could be modified as the form in Eq. (\ref{LIV-n-order})
\beqa
E_p^2 &=&m_p^2+|\vec{p}|^2\left[1 -\eta_n \left(\frac{|\vec{p}|}{E_\text{LV,n}}\right)^n\right] \\
&=&m_p^2+|\vec{p}|^2\left(1 -\eta \frac{|\vec{p}|}{E_\text{LV}}\right).
\label{LIV-1-proton}
\eeqa
Here we choose $n=1$, $\eta_1=\eta$ and $E_\text{LV,1}=E_\text{LV}$.

\hspace{0.4cm}
The following relations are satisfied for determining of the energy threshold of proton~\cite{0211466-thrs-theorem}:
\begin{equation}
\label{b5}
\vec{p}_{N}+\vec{p}_{\pi}=0;  \\ ~~~(p_{r}+p_{p})^2 = (m_{\pi} + m_{N})^2,
\end{equation}
so we can obtain the photon energy at the threshold,
\beqa \nn 
\epsilon^\text{LV}_\text{b(th)} &=& \frac{(m_{\pi} + m_{N})^2 - m_{p}^2}{4E_{p}} + \frac{\eta_\text{n}|\vec{p}|^\text{n+2}}{4E_{p}E_\text{LV,n}^{n}}
\\ &=& \frac{(m_{\pi} + m_{N})^2 - m_{p}^2}{4E_{p}} + \frac{\eta |\vec{p}|^3}{4E_{p}E_\text{LV}},~~(n=1) \nn 
\\ &=&  \frac{(m_{\pi} + m_{N})^2 - m_{p}^2}{4E_{p}} + \frac{\eta E_{p}^2}{4E_\text{LV}},~~(E_p\simeq |\vec{p}|),
\label{modi-eps}
\\
&\to&\frac{(m_{\pi} + m_{N})^2 - m_{p}^2}{4E_{p}}~,~~~~{\rm for}  ~~E_\text{LV}\to\infty, 
\eeqa
where $E_p\simeq |\vec{p}|$ at leading order.
Such a result agrees with the case without Lorentz violation  in Eq.(\ref{org-eps}) when $E_\text{LV}\to\infty$.

\hspace{0.4cm}
We can find the global minimum for $\epsilon^\text{LV}_\text{b(th)}$ from Eq.(\ref{modi-eps}) 
by requiring 
\beqa \frac{\partial \epsilon^\text{LV}_\text{b(th)}}{\partial E_p} = 0~.
\label{eps_min}
\eeqa
The critical proton energy, $E_p = E_{p(\text{cr})}$, which corresponds 
to the minimum of $\epsilon^\text{LV}_\text{b(th)}$, is given by  
\beqa
E_{p(\text{cr})}=\left(\frac{ \left[(m_{\pi} + m_{N})^2 - m_{p}^2\right]E_\text{LV}}{2\eta}\right)^{\frac{1}{3}}.
\label{ep-cr}
\eeqa
\hspace{0.4cm}
It is evident that the critical proton energy $E_{p(\text{cr})}$ increases with $E_\text{LV}$.  Therefore, a higher LIV scale results in a higher critical proton energy.
The two terms in Eq.(\ref{modi-eps}) contribute equally to the threshold photon energy 
$\epsilon^\text{LV}_\text{b(th)}$ when $E_{p}$ takes value at the critical value $E_{p(\text{cr})}$. 
Since the threshold energy $\epsilon^\text{LV}_\text{b(th)}$ in Eq.(\ref{modi-eps}) increases 
with the increment of proton energy $E_p$ upon the critical energy $E_{p(\text{cr})}$,
the numbers of CMB photons that can interact (via phtopion process) with such energetic protons decrease rapidly. 
Therefore, LIV effects could lead to a "reemergence" of ultra energetic protons in the astrophysical spectra.
Such a pattern in the UHE proton spectrum could potentially explain the observations of cosmic rays 
with energies exceeding the traditional GZK cutoff. Observations of ultra high energy  protons (of energy $E_\text{ob}$) may be a sign of LIV upon $E_\text{ob}^3/(m_\pi m_N)$ .

\hspace{0.4cm}
In Eq.(\ref{ep-cr}), when $\eta = -1$, the solution will not be real, 
forcing us to consider the case with $\eta = +1$. 
This choice is opposite to the case of photons, as depicted in Eq.(3) of~\cite{2105.07967,HeMa}.
The reason lies in that photons lack rest mass while protons possess rest masses.
In photopion process $p+\gamma\to \pi+N$, the total rest mass of the final states is larger than that of the initial states.

\hspace{0.4cm}
In the following discussions, we will take $\eta = +1$. The linear order ($n=1$) LIV scale (for photon sector) are determined to be larger than $E_\text{(LV,sub)}= 3.6 \times 10^{17}$ GeV
for subluminal photon ($\eta=-1$) from Gamma Ray Burst (GRB) constraints~\cite{1607.03203,1607.08043,1801.08084,2108.05804,2105.07967}. If we naively adopt such a characteristic LIV scale for the proton LIV case,  we can obtain the critical proton energy $E_{p(\text{cr})}$ 
\beq
E_{p(\text{cr})} \simeq 1.723\times 10^5 ~{\rm GeV},
\label{ep_cr}
\eeq
with the corresponding lower threshold energy scale for CMB photon 
\beq
\epsilon^\text{LV}_\text{b(th)}|_\text{min} \approx 4 \times 10^2 ~{\rm eV}.
\eeq
Such a value is almost six orders of magnitude greater than the average energy of CMB photon, 
making it almost impossible for any CMB photon to reach this energy threshold.

\hspace{0.4cm}
However, the value of $E_\text{LV}$ can differ significantly in the proton LIV sector compared to that in the photon LIV sector.
There are many discussions on the bounds for $E_\text{LV}$~\cite{1607.03203,1607.08043,1801.08084,2108.05804,2105.07967,2105.06647}, among which
the strongest one is (effectively) $E_\text{LV}^\text{eff}\simeq10^{29}$ GeV~\cite{1810.13215} (which, in fact, corresponds to very small LIV effect with $\eta\simeq 10^{-10}$ for $E_\text{LV}= M_\text{Pl}$). 
 Although an effective transPlanckian scale may seem unnatural, it could indicate the presence of some UV mechanism that strongly suppresses the LIV effect, resulting in a very small value of $\eta$. Therefore, we adopt $E_\text{LV}^\text{eff} \simeq 10^{29}$ GeV as the maximum value for LIV, as this value remains consistent with all phenomenological constraints \cite{1607.03203,1607.08043,1801.08084,2108.05804,2105.07967}.

 \hspace{0.4cm}
In Fig. \ref{fig2:epsilon-elv}, we show the values of $\epsilon^\text{LV}_\text{b(th)}$ versus $E_\text{LV}$ for $E_\text{LV}$ ranging from $10^{17}$ GeV to $10^{29}$ GeV. It is evident from this figure that the minimum energy threshold for CMB photons required to allow their interaction with UHECR protons via photopion processes is significantly higher (by several orders of magnitude) than the average energy of CMB photons, which is approximately $2.6 \times 10^{-4}$ eV. Consequently, with the presence of LIV in the proton sector, CMB photons become significantly less effective at attenuating ultra-energetic protons through photopion processes once the proton energy exceeds the critical value given in Eq.(\ref{ep_cr}).

%
\begin{figure}[htb]
\centering
\includegraphics[scale=0.62]{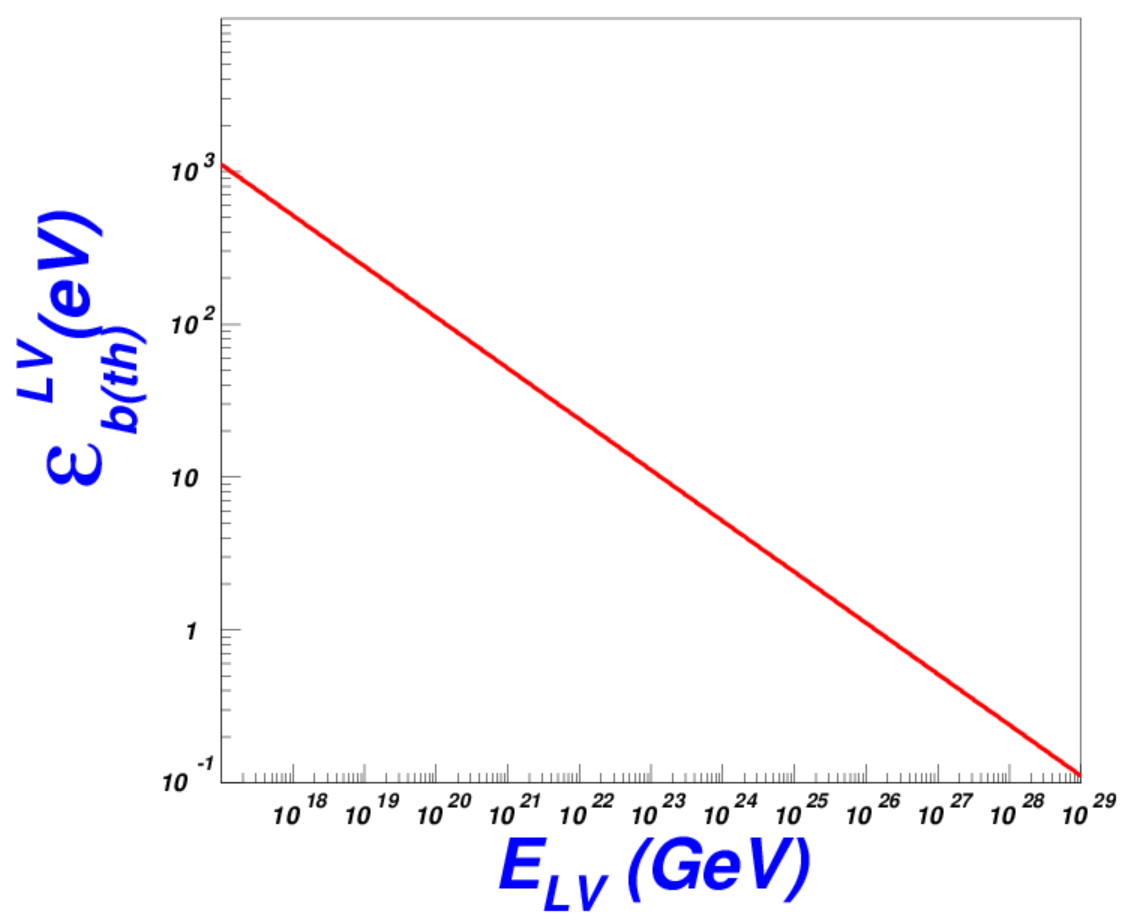}
\caption{$\epsilon^\text{LV}_\text{b(th)}$ varies with $E_\text{LV}$.}
\label{fig2:epsilon-elv}
\end{figure}

\hspace{0.4cm}
Note that the critical threshold $E_{p(\text{cr})}$ in Eq.(\ref{modi-eps}) increases dramatically with the increment of the proton energy $E_p$.
If we adopt the upper range of $E_\text{LV}$ at around $E^\text{eff}_\text{LV}\sim 10^{29}$ GeV, 
we obtain  $\epsilon^\text{LV}_\text{b(th)}\simeq 0.1$ eV, which is significantly larger than the average energy of CMB photons. The critical proton energy for $E^\text{eff}_\text{LV}\sim 10^{29}$  can be determined to be $10^{18}$ eV, which is significantly smaller than the event observed by Akeno-AGASA experiment~\cite{AGASA-1,AGASA-2} that have an  energy at almost $E_p(max) = 10^{21}$ eV. Thus, protons in UHECRs at energies on the order of $10^{21}$ eV could potentially reach Earth from distant sources without experiencing significant attenuation by CMB photons during their propagation. Consequently, such beyond-GZK-cutoff events could potentially be detectable \cite{AGASA-1,AGASA-2,uhecr-6,Hires-2,pao-1,pao-2,TA-1}.

\begin{figure}[htb]
\centering
\includegraphics[scale=0.7]{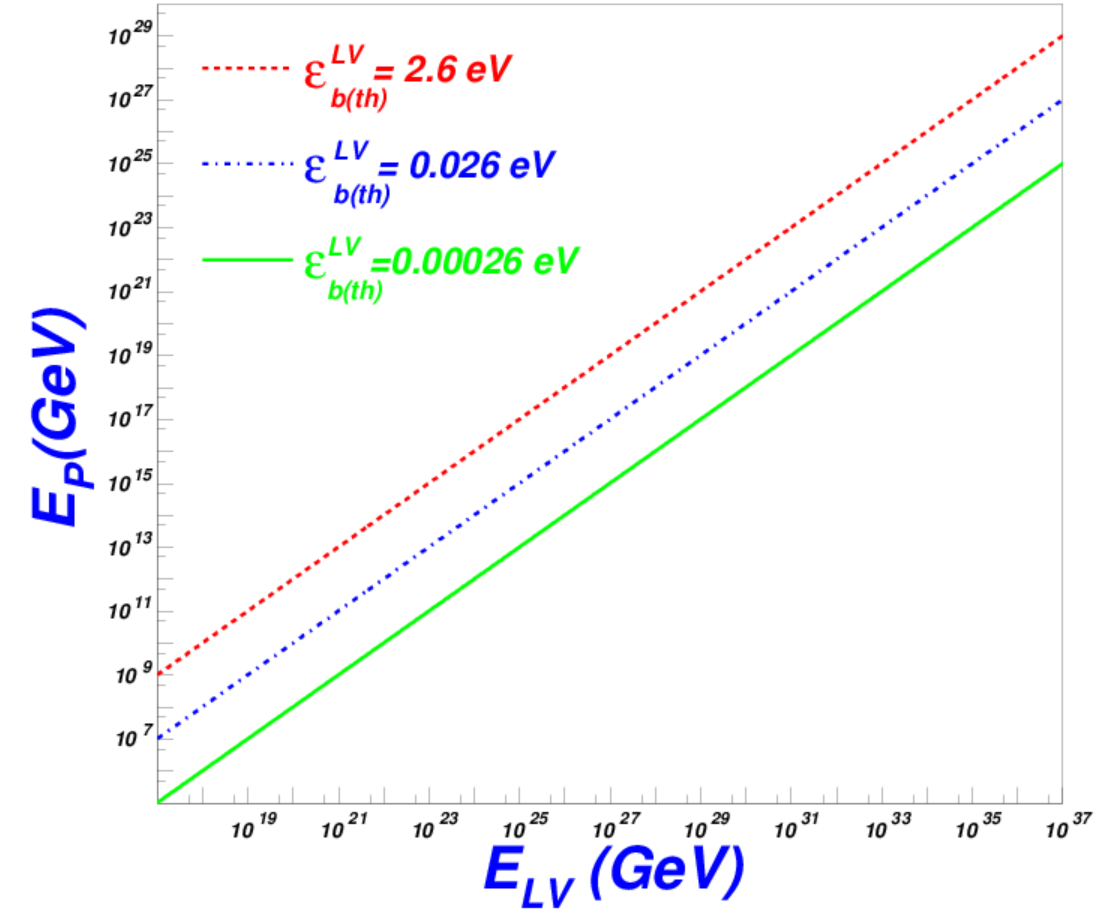}
\caption{$E_{p}$ v.s. $E_\text{LV}$ for $\epsilon^\text{LV}_\text{b(th)}=2.6\times 10^{-4}$ eV, $2.6\times 10^{-2}$ eV, $2.6\times 10^{0}$ eV, respectively.}
\label{fig3:ep-elv}
\end{figure}

\hspace{0.4cm}
We show in Fig.\ref{fig3:ep-elv} the allowed values of proton energy $E_p$ versus the LIV scale $E_\text{LV}$ when threshold energies for CMB photons required for photopion processes are fixed.
It can be seen that, given the value of $\epsilon^\text{LV}_\text{b(th)}$, higher energies of cosmic ray protons are required for higher LIV violation scale $E_\text{LV}$.

\hspace{0.4cm}
Observed UHECR data can be used to impose stringent constraints on the low energy effective LIV theory, which may emerge as an effective description of some UV extensions of the SM~\cite{9703464-sme-cpt-sm,9809521-Koste1,sme-3}. 
For example, with the observation of energetic cosmic ray protons around $10^{20}$ eV, one can constrain the LIV coefficient $\eta$ to be approximately $\eta\sim 10^{-10}$ if  the LIV scale is assumed to lie near the Planck scale. Such an $\eta$ value is significantly lower than its natural choice $\eta\sim {\cal O}(1)$ that one might anticipate. This a stringent constraint imply that the UHECR data could shed new light on the test of LIV theories.

\item  Both $n=1$ and $n=2$ order LIV terms are present:

\hspace{0.4cm}
In the case where both $n=1$ and $n=2$ order LIV terms are present, the threshold energy for the photons is given as 
\beqa
\epsilon^\text{LV}_\text{b(th)}&=& \frac{(m_{\pi} + m_{N})^2 - m_{p}^2}{4E_{p}} + \frac{\eta_{p;\text{L}} |\vec{p}|^3}{4E_{p}E_\text{LV}}+\frac{\eta_{p;\text{NL}}|\vec{p}|^{4}}{4E_{p}E_\text{LV}^2}, 
\\ &=&  \frac{(m_{\pi} + m_{N})^2 - m_{p}^2}{4E_{p}} + \frac{\eta_{p;\text{L}} E_p^2}{4 E_\text{LV}}+\frac{\eta_{p;\text{NL}} E_p^{3}}{4 E_\text{LV}^2}, 
\label{modi-eps:3}
\eeqa
\hspace{0.4cm}
The extremum for $\epsilon^\text{LV}_\text{b(th)}$ can be determined from Eq.(\ref{modi-eps:3}) through Eq.(\ref{eps_min}),
resulting in the quartic equation
\beqa
3 \eta_{p;\text{NL}} E_p^4+2 \eta_{p;\text{L}} E_p^3E_\text{LV}-\left[(m_{\pi} + m_{N})^2 - m_{p}^2\right]E_\text{LV}^2=0~.
\label{quartic:root}
\eeqa
\hspace{0.4cm}
Although an quartic equation can be solved analytically, the form of its roots are rather complicated. Therefore, it is more practical to resort to numerical methods for its solutions. For very  large $E_\text{LV}$, we can always find an approximate root with $E_p\approx -\frac{2\eta_{p;\text{L}}}{3\eta_\text{p:NL}} E_\text{LV}$. Therefore, from Vieta's theorem, we anticipate that the remaining three solutions have magnitude of the order $E_{p(\text{cr})}\equiv(m_{\pi}m_{N}E_\text{LV})^{1/3}$, which is the same order as Eq.(\ref{ep-cr}). However, for certain fixed choices of $E_\text{LV}$, it is possible the values of  $\epsilon^\text{LV}_\text{b(th)}$ with respect to $E_p$ exhibit multiple local minima and maxima. If the value of typical local maximum exceeds the average CMB photon energy, the GZK cutoff in this case displays a discontinous pattern. That is, protons with energies in several disconnected ranges can survive the GZK constraints. The existence of such a pattern requires that Eq.(\ref{modi-eps:3}) should have three local minima/maxima with one saddle point, which corresponds to the requirement that Eq.(\ref{quartic:root}) should have four distinct real roots. Observations of such discontinuous GZK cutoff bands can be a sign of this scenario. Additionally, contrary to naive expectations, the presence of higher-order LIV terms can play an important role in determining the GZK cutoff energy bands.    
\end{itemize}

\hspace{0.4cm}
We aim to analyze the root structure of the quartic equation. The numbers of real roots of Eq.(\ref{quartic:root}) determine the local minima of $\epsilon^\text{LV}_\text{b(th)}$. On the other hand, the real roots are determined by the discriminant of the quartic equation. For a quartic equation of the form
\beqa
x^4+q x^2+r x+s=0~,
\label{standard:quartic}
\eeqa
the discriminant is given by
\beqa
\Delta=16 q^4 s-4 q^3 r^2-128 q^2 s^2+144 q r^2 s-27 r^4+256 s^3~.
\eeqa
An necessary and sufficient condition for the existence of four distinct real roots requires that:
\beqa
\Delta>0,~q<0,~s>\f{q^2}{4}.
\label{necessary}
\eeqa

In our case, the equation Eq.(\ref{quartic:root}) can be reformulated into the form of Eq.(\ref{standard:quartic}) after dividing by the coefficient $3\eta_{p;\text{NL}}$ and replacing 
\beqa
E_p\rightarrow 
\tl{E}_p
=E_p+\f{\eta_{p;\text{L}}E_\text{LV}}{6\eta_{p;\text{NL}}}.
\eeqa
The quartic equation Eq. (\ref{quartic:root}) reduces to 
\beqa
\tl{E}_P^4+ q \tl{E}_p^2+r \tl{E}_P+s=0~,
\label{true:quartic}
\eeqa
with
\beqa
q&=&-\f{1}{6} \(\f{ \eta_{p;\text{L}} E_\text{LV}}{\eta_{p;\text{NL}}}\)^2~,
~~~~~r=\f{1}{27}\(\f{ \eta_{p;\text{L}} E_\text{LV}}{\eta_{p;\text{NL}}}\)^3~,\nn\\
s&=&-\f{\left[(m_{\pi} + m_{N})^2 - m_{p}^2\right]E_\text{LV}^2}{3 \eta_{p;\text{NL}}}-\f{1}{432}\(\f{ \eta_{p;\text{L}} E_\text{LV}}{\eta_{p;\text{NL}}}\)^4~.
\eeqa
As physical energy $E_p$ should take positive values, the four real roots are required to be positive. Therefore, we impose the following additional constraints  
\beqa
\eta_{p;\text{NL}} f(E_p)>0, ~~~{\rm for} ~~~~~~E_P<0,
\label{positive:E_p}
\eeqa
in addition to Eq.(\ref{necessary}), with 
\beqa
f(E_p)\equiv 3 \eta_{p;\text{NL}} E_p^4+2 \eta_{p;\text{L}} E_p^3E_\text{LV}-\left[(m_{\pi} + m_{N})^2 - m_{p}^2\right]E_\text{LV}^2~.
\eeqa

The existence of discontinuous GZK cutoff energy bands
 requires that the above Eq.(\ref{true:quartic}) satisfies the conditions in Eq.(\ref{necessary}) and Eq.(\ref{positive:E_p}).

\item LIV in both the proton  and photon sectors:

\hspace{0.4cm}
If LIV originates from the underlying spacetime structures, dispersion relations for gauge bosons, such as photons, should also be modified to incorporate the LIV effects. Therefore, for completeness, we will also include the LIV effects for initial state photons. 

\hspace{0.4cm}
 The MDR for photons can take the following form as\cite{0510172,1406.4568,2105.07967}
\beq
E_\gamma^2=|\vec{p}|^2\left[1-\sum\limits_{n}\eta_{\gamma;n}
\left(\frac{|\vec{p}|}{E_\text{LV}}\right)^n\right].
\eeq
Therefore, with LIV effects for both proton and photon sectors, the threshold energy for the photons can be determined to be
\beqa 
\epsilon^\text{LV}_\text{b(th)} &=& \frac{(m_{\pi} + m_{N})^2 - m_{p}^2}{4E_{p}} + \frac{\eta_{p;n_p}|\vec{p}|^{n_p+2}}{4E_{p}E_{p;\text{LV}}^{n_p}}+\frac{\eta_{\gamma;n_\gamma}|\vec{p}_\gamma|^{n_\gamma+2}}{4E_{p}E_{\gamma;\text{LV}}^{n_\gamma}},\nn \\
&=&\frac{(m_\pi + m_N)^2 - m_p^2}{4E_p} + \frac{\eta_{p;n_p}|\vec{p}|^{n_p+2}}{4E_{p}E_{p;\text{LV}}^{n_p}}+\frac{\eta_{\gamma;n_\gamma}(\epsilon^{LV}_{b(th)})^{n_\gamma+2}}{4E_{p}E_{\gamma;\text{LV}}^{n_\gamma}}.
\eeqa
\hspace{0.4cm}
For simplification, we focus on the case with $n_p=1$ and $n_\gamma=1$ in our following discussions. 

\hspace{0.4cm}
The LIV scales for both sectors could be hierarchical,  
$$E_{{LV}}\equiv E_{{p;\text{LV}}}\gg E_{{\gamma;\text{LV}}}.$$
Thus, we obtain the threshold energy for the photon 
\beqa
\epsilon^\text{LV}_{b(th)}&=& \frac{(m_{\pi} + m_{N})^2 - m_{p}^2}{4E_{p}} + \frac{\eta_p |\vec{p}|^3}{4E_{p}E_{p;\text{LV}}}+\frac{\eta_{\gamma}|\vec{p}_\gamma|^{3}}{4E_{p}E_{\gamma;\text{LV}}}, \nn 
\\ &=&  \frac{(m_{\pi} + m_{N})^2 - m_p^2}{4E_p} + \eta_p\frac{ E_{p}^2}{4E_\text{LV}}+\frac{\eta_{\gamma}(\epsilon^\text{LV}_{b(th)})^{3}}{4E_pE_{\gamma;\text{LV}}}, 
\label{modi-eps:2}
\eeqa
with $E_p\simeq |\vec{p}|, \epsilon^\text{LV}_\text{b(th)}\simeq|\vec{p}_\gamma|$ when LIV effects are present in both proton and photon sectors. The second term in Eq.(\ref{modi-eps:2}) dominates the  contribution to $\epsilon^\text{LV}_\text{b(th)}$ when $E_p$ is very large, satisfying
$$E_p \gg {E}_{p(\text{cr})}\simeq\left[ m_\pi m_N E_\text{LV} \right]^{1/3}. $$

\hspace{0.4cm}
So we have
\beqa
\epsilon^\text{LV}_\text{b(th)} &\approx&  \frac{(m_{\pi} + m_{N})^2 - m_{p}^2}{4E_{p}} + \eta_p\frac{ E_{p}^2}{4E_\text{LV}}+\frac{\eta_{\gamma}\eta_p^3E_p^5}{32 E_\text{LV}^3 E_{\gamma;\text{LV}}}, 
\label{modi-eps:22}
\eeqa
in the large $E_p$ regions. We can see that the third term can be non-negligible only if $E_\text{LV}\gtrsim E_p\gg E_{\gamma;\text{LV}}$. Current lower bounds on $E^{n_\gamma}_{\gamma,\text{LV}}$ are rather stringent, with constraints requiring it to be much higher than
the Planck scale for $n_\gamma=1$. Thus, the presence of LIV in the photon sector with $n_p=1$ and $n_\gamma=1$ will not lead to additional non-negligible contributions to the threshold energy $\epsilon^\text{LV}_{b(th)}$, whose value remains almost the same as the case without LIV in the photon sector. 

\hspace{0.4cm}
If the leading LIV effect for photons is of quadratic order ($n_\gamma=2$), by similar reasoning as in the case $n_\gamma=2$ and $n_p=1$, one can conclude again that the quadratic LIV term of the photon sector can lead to non-negligible contribution to the threshold energy of photon only if $E_\text{LV}\gtrsim E_p\gg E^{n_\gamma=2}_{\gamma;\text{LV}}$. In this case, the bound on $E^{n_\gamma=2}_{\gamma,\text{LV}}$ is less stringent, and could possibly be smaller than the Planck scale. Therefore, LIV in the photon sector could increase the photon threshold energy $E^{n_\gamma}_{\gamma,\text{LV}}$ for a given $E_P$ when $E_\text{LV}\gtrsim E_p\gg E^{n_\gamma=2}_{\gamma;\text{LV}}$, making it easier for the UHECR protons to surpass the GZK cutoff scale. This conclusion also holds for cases where the leading LIV effect for photons involves
 $n_\gamma\geq 2$.  




\end{itemize}



\section{Conclusions}\label{sec:conclusions}

Tiny Lorentz Invariance Violation (LIV) effects may arise from specific space-time structures predicted by quantum gravity theories. Consequently, it is reasonable to expect that such tiny LIV effects could manifest in the proton sector.
We find that, in the presence of these small LIV effects, the threshold energy for photons to initiate photopion interactions with protons can be significantly increased—ranging from approximately 0.1 eV to $10^3$ eV—compared to the scenario without LIV.
This implies that protons in Ultra-High-Energy Cosmic Rays (UHECRs) could potentially travel much greater distances without significant energy loss due to photopion interactions with CMB photons.
This mechanism may provide an explanation for the observed beyond-GZK cutoff events. Furthermore, we find that when both leading-order and next-to-leading-order LIV effects are considered, the higher-order LIV terms can introduce discontinuous GZK cutoff energy bands.
To the best of our knowledge, no prior discussions have addressed the possibility of discontinuous GZK cutoffs. Observations of beyond-GZK cutoff UHECR events involving protons could thus serve as a powerful tool to constrain the scale of LIV.
These UHECR events offer an exceptionally sensitive probe for investigating LIV effects and could provide new insights into ultraviolet (UV) LIV theories near the Planck scale.

The form of the LIV term in the proton sector also maintains spatial-temporal isotropy. As a result, correlations between the anisotropy in UHECR arrival directions and the spatial distributions of potential astrophysical sources could be used to further constrain the relevant LIV parameters. We plan to explore these aspects in future studies.

\section*{Acknowledgments:}
This work was supported by the National Natural Science Foundation of China (NNSFC) under grant Nos.12075213 and 12335005, and by the Natural Science Foundation for Distinguished Young Scholars of Henan Province under grant number 242300421046.



\end{document}